# Interfacial charge transfer influences thin-film polymorphism

*Fabio Calcinelli[1], Andreas Jeindl[1], Lukas Hörmann[1], Simiam Ghan[2], Harald Oberhofer[2] and Oliver T. Hofmann[1]\**

[1]Institute of Solid State Physics, Graz University of Technology, 8010 Graz, Austria

[2]Chair for Theoretical Chemistry and Catalysis Research Center, Technical University Munich, 85748 Garching, Germany

**\*Corresponding Author: o.hofmann@tugraz.at**

ABSTRACT:
The structure and chemical composition are the key parameters influencing the properties of organic thin films deposited on inorganic substrates. Such films often display structures that substantially differ from the bulk, and the substrate has a relevant influence on their polymorphism. In this work, we illuminate the role of the substrate by studying its influence for para-benzoquinone on two different substrates, Ag(111) and graphene. We employ a combination of first principles calculations and machine learning to identify the energetically most favorable structures on both substrates and study their electronic properties.

Our results indicate that for the first layer, similar structures are favorable for both substrates. For the second layer we find two significantly different structures. Interestingly, graphene favors the one with less, while Ag favors the one with more electronic coupling. We explain this switch in stability as an effect of the different charge transfer on the two substrates.

Organic thin films are materials of increasing interest, mainly by virtue of their application to the field of organic electronics. In comparison to inorganic alternatives, they present advantages such as mechanical flexibility and low cost. With a thickness ranging from less than a nanometer up to a few micrometers, organic thin films are commonly employed in the construction of Organic Field-Effect Transistors (OFET),[1-2] Organic Light Emitting Diodes (OLED)[3] and Organic Solar Cells.[4] Of particular interest are films composed of molecules that form ordered films with relatively high charge-carrier mobilities.[5–7] In fact, the properties of molecular materials, and especially their charge-carrier mobilities, depend drastically on the polymorph they assume, i.e. the relative arrangement of individual molecules in the thin film.[8,9]

Which polymorph a thin film forms depends not only on the fabrication conditions,[10] but also the nature of the substrate on which it grows has a decisive impact. Because the substrate interacts with molecules in the first layer, and because it changes the way molecules interact with each other (e.g., because they become charged), the second and subsequent layers can either assume the same structure as the first,[11–13] assume a bulk structure,[14] or form a completely different structure altogether.[15] The decisive role of the substrate is highlighted by reports where even the same molecule forms different structures on different substrates.[16–18]

In this work, we shine light on the role of the substrate and tackle the question whether – and why – some substrates are more likely to induce polymorphs which are beneficial for organic electronics than others. To that aim, we use a combination of machine learning and first-principles calculations to investigate the structure of thin films of para-benzoquinone adsorbed on Ag(111) and on graphene.

Both substrates are sensible electrode materials in organic electronics.[19–21] At the same time, they show fundamentally different interactions with organic molecules: Ag is a weakly reactive substrate, which readily undergoes charge-transfer reactions and can form weak covalent bonds with organic adsorbates.[22–29] Conversely, graphene hardly forms covalent bonds at all. Benzoquinone was chosen as model molecule due its small size (reducing the computational cost) while exhibiting π-conjugation and functionalization with carbonyl groups. As we have previously shown, the intermolecular interactions of this molecule are qualitatively similar to those of technologically more relevant, larger analogues, like 5,12-pentacenequinone.[29–32]

Predicting the polymorphism of thin films is, however, far from trivial. To date, a variety of specialized algorithms are available which predict the structures of molecular crystals[33–41], of their surfaces,[42] of single molecules adsorbing on a surface,[43–47] or monolayers of molecules adsorbed on a substrate.[48–53]

Here, we use an extended version of the SAMPLE approach, which is specifically designed for inorganic/organic interfaces.[51] When applying the SAMPLE approach, one starts with finding the local adsorption geometries that an isolated molecule could adopt on a surface. These structures act as building blocks. As a second step, possible polymorphs – with numbers ranging in the millions – are built by assembling all possible combinations of these building blocks in a variety of unit cells. A small subset of these polymorphs is then evaluated with first-principle calculations (i.e., dispersion-corrected DFT, see Method section), and the resulting energies are used to train an energy model utilizing Bayesian linear regression. The trained energy model can then predict the energies of all remaining polymorphs with a level of accuracy similar to the underlying electronic structure method. To predict the second layer, we employ the SAMPLE approach a second time, now taking a suitable first adlayer as a new substrate (see below). A more detailed explanation of the SAMPLE procedure is given in ref [51].

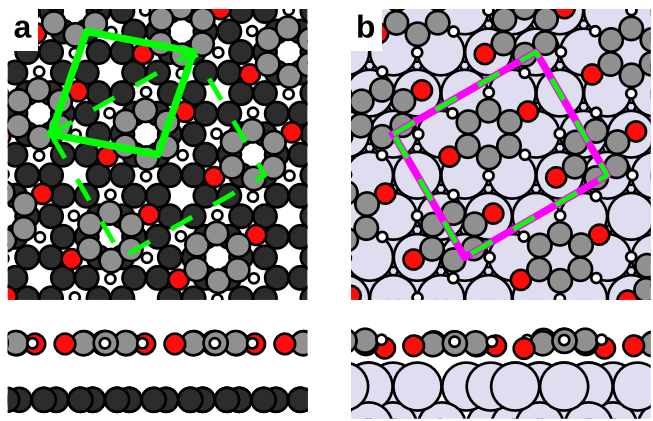

*Figure 1. Geometry of the first layer of benzoquinone on (a) graphene and (b) Ag(111). The unit cell for benzoquinone on graphene is shown in solid green, and the unit cell for benzoquinone on Ag(111) is shown in purple. The dashed green lines indicate a unit cell equivalent to the unit cell on graphene and twice as large (its (1, 1, -1, 1) transform), which fits the Ag unit cell (purple) almost perfectly.*

Before considering thin-film growth it is necessary to look at the structure that the first layer of benzoquinone forms on the two substrates. For Ag(111) the polymorph candidates have been obtained in an earlier work,[29] while for graphene a structure search is performed anew through the SAMPLE approach (see Supporting Information). The best polymorph in the SAMPLE ranking has one molecule per unit cell and is presented in Figure 1a. In this geometry molecules adsorb at a height of approximately 3.3 Å and remain almost perfectly flat.

For benzoquinone on Ag(111), we find a comparable structure among the energetically best polymorph candidates (details in the Supporting Information). This configuration is shown in Figure 1b. Its unit cell contains two molecules, placed on a top site and on a bridge site of the metal surface. The molecules adsorb at a height of about 2.6 Å, and are slightly bent, with the oxygen closer to the metal substrate than the carbon backbone.

The two geometries appear strikingly similar, and in fact an equivalent cell of the graphene monolayer, with twice the area, is virtually identical (deviations lower than 0.01 Å) to the cell of the monolayer on Ag(111) (dashed green cell and purple cell in Figure 1b). The fact that the first layer on both substrates shows equivalent lattice parameters and molecular alignment means that any subsequent layers will be subjected to identical stress, and to equivalent templating effects from the first layer. In other words, we can expect that any differences in the energetics and structure of the second layer stems directly from the (electronic) influence of the substrate.

As a first step in describing thin films, we study the second molecular layer, and we invoke two assumptions. First, we assume that the geometry of the first layer only undergoes minor changes when additional material is deposited – in particular, that the unit cell remains fixed. We note that, in practice, this is not always the case, as in some systems the first layer re-orients to form a more tightly packed layer.[54–56] However, predicting such re-orientations is beyond the scope of the present work. Second, we assume Frank-Van Der Merwe growth, i.e. each layer does not start forming until the previous layer is full. These two assumptions allow us to use the SAMPLE approach. For this we employ the monolayer geometries of benzoquinone (plus metal/graphene) as effective substrate unit cells and search for and combine the local adsorption geometries in the second layer. In the case of graphene, the search for second-layer single molecule adsorption geometries was conducted directly on the full system (substrate + first layer), while for Ag, because of the large amount of required computational resources, they were conducted on the Ag(111)-monolayer of benzoquinone without metal atoms. In order to obtain accurate energies, after the ranking of the polymorphs candidates by SAMPLE we perform full geometry optimizations for the 10 best structures (see Methods section). During these optimizations, the energies still change notably (see Supporting Information), as the molecules in the second layer assume more favorable orientations towards the first layer.

For Ag, the five energetically best bilayer structures are shown in Figure 2a. The ranking is performed according to energy per area, the most sensible measure for the stability of closed-packed adsorbate polymorphs.[57] The full prediction data for both systems, including a comparison between predicted and calculated energies, can be found in the Supporting Information. In the energetically most favorable structure, the benzoquinone molecules in the first and the second layer are partly on top of each other, with the (negatively charged) oxygen of one molecule always aligned with the center (i.e., the least negative region) of the ring of the molecules in the other layer. We refer to this alignment, that is shown in Figure 2a in by red molecules in the top layer, as molecule-on-molecule (MoM) hereafter. The second-best geometry is already 50 meV/nm² worse in energy. In this geometry the molecules in the second layer are located above "gaps" of the first layer (marked in orange in Figure 2a). Only the carbonyl-groups of the first and the second layer are on top of each other, with oppositely directed dipoles presumably leading to electrostatic attraction. To distinguish this alignment from the others, we refer to it as molecule-on-gap (MoG) hereafter. The energetically next-higher lying structures are combinations of MoM and MoG, variations thereof, and structures with lower coverages.

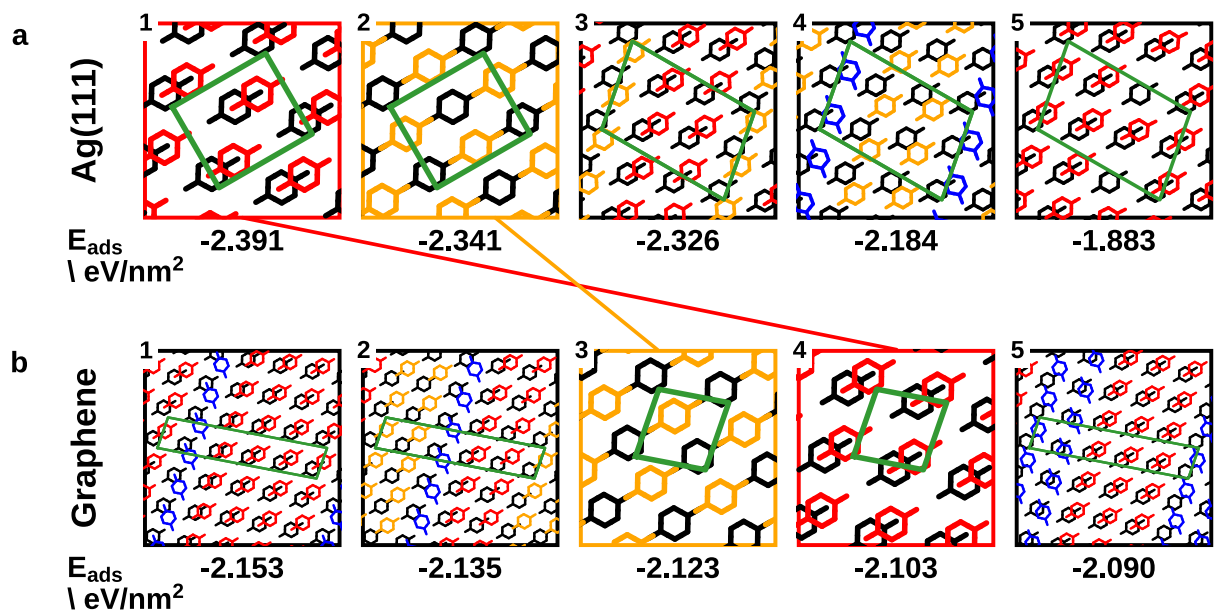

*Figure 2. Adsorption energy and graphical representation of the five best configurations of the bilayer of benzoquinone on Ag(111) and graphene. The boxes corresponding to Molecule-on-Gap and Molecule-on-Molecule (for explanation see main text) are colored in orange and red, respectively. In the geometry representations Ag and graphene are omitted, the first layer of adsorbates is colored in black, and the second layer is colored according to the adsorption positions (i.e. similar positions have the same color).*

On graphene, we also find the MoM and the MoG geometry as energetically favorable structures. However, in salient contrast to the situation on Ag, here the MoG structure is energetically more beneficial than MoM by 20 meV/nm². Only two structures are found that are energetically even better than MoG and MoM. Both of these structures are noticeably more complex than MoG and MoM, featuring five adsorbates per unit cell and several adsorption positions similar to MoM and MoG. For the sake of conciseness and clarity, we will focus the following discussion on the MoM and the MoG structures only. A brief discussion of structures 1 and 2 can be found in the Supporting Information.

Since the charge-carrier mobility (or, more precisely, the electronic coupling) of a crystal depends on the wave-function overlap, [8,58,59] already a visual inspection of the MoM and MoG geometries lets us expect that this property will be very different for the two geometries. The fact that the ordering of the two polymorphs reverses depending on the substrate, therefore, deserves further scrutiny, and we should attempt to explain the reasons of this switch and its consequences on interlayer electronic coupling.

When considering only the second layer, on each substrate the MoM and MoG polymorphs exhibit the same unit cell vectors and very similar geometries, differing mostly by a translation relative to the first benzoquinone layer. Thus, we expect the switch in the energetic ordering to be caused by a variation in the interlayer interactions between the first and the second layer.

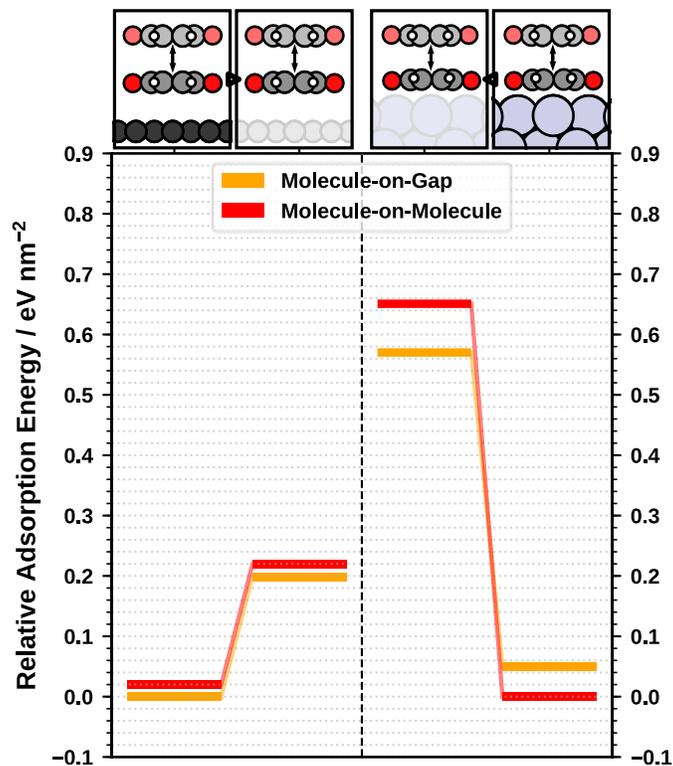

*Figure 3. Adsorption energies of MoM and MoG adsorbing on the two monolayer-on-substrate geometries (left: graphene; right: Ag(111)), and on gasphase monolayers having the same geometry as the adsorbed monolayers, but with no substrate atoms. The energies are given relative to the value of the most stable geometry for each full-substrate system.*

To verify that the switch in stability is caused directly by the different substrates, and not by the small geometric differences in the first layer, we examine the variation in adsorption energy that occurs if we keep the geometry of the first layer fixed, but remove all graphene or Ag atoms. We find that for the case of graphene, MoM and MoG suffer destabilitazions that are modest and fundamentally equivalent, i.e. a graphene substrate does not notably affect the energetic ordering. For Ag(111), when removing the substrate MoM becomes energetically destabilized with respect to the MoG geometry. This indicates a stronger influence of the substrate on the MoM structure compared to MoG. We can thus conclude that the Ag substrate massively changes the way the first and the second layer interact with each other. Specifically, we find that the Ag substrate significantly stabilizes the MoM geometry, explaining why it is favored on Ag, but not on graphene.

We now need to ask which underlying mechanism stabilizes the MoM geometries. We can trace the effect back to the charge rearrangements resulting from the contact between the substrate and a molecular layer. To illustrate this, we calculated the adsorption-induced charge rearrangements $\Delta\rho$, defined as

$$\Delta\rho = \rho_{system} - \rho_{sub} - \rho_{monolayer} \quad \textit{Equation 1}$$

where $\rho_{system}$, $\rho_{sub}$ and $\rho_{monolayer}$ are the charge densities of the combined system, of the substrate and of the isolated benzoquinone monolayer respectively. We further calculate the net charge transfer by estimating the maximum value of:

$$Q_{bond}(z) = \int_0^z \Delta\rho(z')dz' \quad \textit{Equation 2}$$

for the benzoquinone monolayers on Ag and on graphene. These calculations lead to a value of $Q_{bond}$ of -0.498 for benzoquinone on Ag(111), and -0.031 for benzoquinone on graphene.

In other words, for Ag, half an electron is transferred from below the substrate surface to above it. Conversely, graphene is practically inert, and the electron transfer is negligible. Furthermore, by conducting a Molecular-Orbital Projected Density of States analysis,[60,61] we find that the LUMO of the benzoquinone layer gets filled in the case of Ag, reaching an occupation of 1.25 electrons, while in the case of graphene it remains substantially empty at an occupation of 0.05 electrons. Details on both the net charge transfer and MODOS calculations can be found in the Supporting Information.

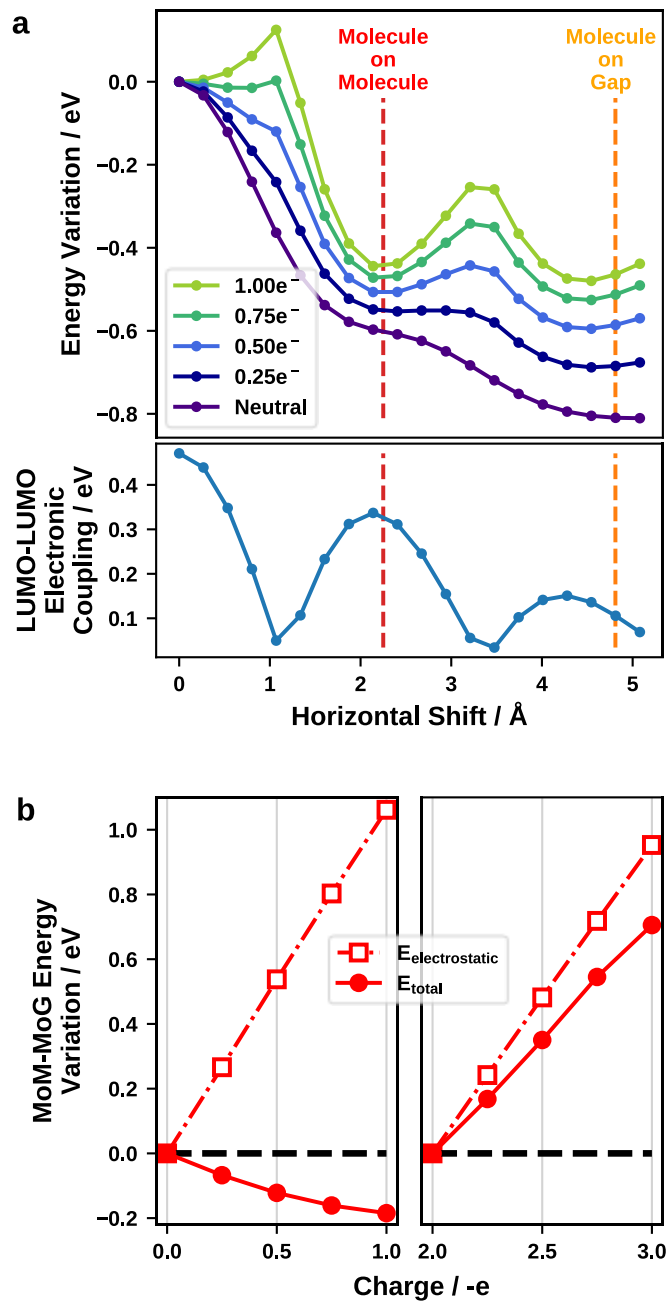

*Figure 4. a) Variation of the total energy (without van der Waals interactions) and LUMO-LUMO electronic coupling for different shifts along the main molecular axis of a benzoquinone dimer. The shifts corresponding to the MoM and MoG structures are indicated with vertical lines. b) Variation of the energy difference between MoM and MoG as a function of extra charge. Here, in addition to the values of charge used in (a), we see the effect of charges in the 2-3 e- range. In this range, additional charge occupies the antibonding orbital combination.*

This different charge transfer directly impacts the interaction with the second molecular layer. To analyze the effect of extra charge on interlayer interaction we use a simple dimer model composed of two stacked benzoquinone molecules. The two molecules are arranged at a distance of 3 Å along the z direction, which is a reasonable approximation of the interlayer distances for our systems. They are then shifted with respect to one another along the long molecular axis. The shifting starts from a position of congruence in x-y coordinates, and includes positions corresponding to both the MoM and MoG offsets. For each position, the electronic energy of the system (i.e. the total energy without van der Waals contributions) is evaluated together with the coupling between the LUMOs of the two molecules (Figure 4a). The suitability of this model for describing the interactions of the full monolayers is discussed in the Supporting Information.

It has been observed that, in analogous cases, one can find an inverse correlation between stability and HOMO-HOMO coupling, as a consequence of Pauli repulsion.[62–64] In our case, though, we are interested in the response of the system to the introduction of additional electronic charge, and therefore we focus on the coupling between LUMOs. For the neutral system (shown in purple), there is no correlation between the coupling and the energy. This also wouldn't be expected, since the orbitals are completely empty. Rather, the energy of the system decreases systematically as the molecules are shifted away from each other. This can be attributed to a reduction in Pauli-Pushback, as the wave-functions no longer overlap.

The situation changes notably when additional charge is introduced. As can be seen, particularly for larger charges, the energy profile now shows an inverse correlation with the LUMO-LUMO coupling, i.e. situations with a large coupling are energetically more favorable than those with a small coupling. The MoM geometry has a significantly larger coupling than the MoG geometry (although both are local maxima) and is therefore more stabilized (up until a charge of two electrons, see below). This is in accordance with what we have observed in the behavior of the configurations in Figure 3.

This behavior can be readily rationalized by valence-bond theory: When two identical molecules come in contact, their LUMOs (originally at the same energy) will hybridize and form a bonding and an antibonding linear combination. The splitting depends on the orbital coupling,[65] i.e. the bonding combination is more strongly bonding the larger the coupling is. If the system is neutral, this has no effect on the total energy. However, when electrons are introduced, they will first occupy the bonding linear combination. As long as there are less than two additional electrons per dimer, only the bonding one will be occupied, resulting in a net energy gain that is larger the larger the coupling is. Conversely, when more than two electrons are introduced, the effect reverses. This tendency is confirmed by Figure 4b, where the variation of the MoM-MoG energy differences is plotted as a function of charge. One can observe that MoM is favored when increasing charge between 0 and 1 electrons but is disfavored when increasing charge between 2 and 3 electrons. For each value of additional charge, a term describing the pure electrostatic interaction between layers has been calculated. This term is obtained by applying an energy decomposition scheme in which the electron densities of the isolated fragments are combined and their classical electrostatic energy is calculated.[64,66–68] One can see that this electrostatic term disfavors MoM for all values of additional charge, proving that the stabilization of MoM in the 0-1 electrons range is caused by the previously discussed orbital hybridization, and not by purely electrostatic effects. In other words, we have demonstrated that the charging of the first layer on Ag(111) is the main factor governing the preferability of MoM compared to MoG, because the additional charge in the first layer directly benefits geometries with a large LUMO-LUMO overlap.

This provides a simple and solid explanation of why the two arrangements present a different stability on the two substrates. In addition, it provides an important hint towards the consequences of this stability switch: it is known that charge carrier mobility, within the model of the hopping regime, is fundamentally influenced by the coupling between the origin and destination orbitals.[59] Generally, our results indicate that substrates that undergo significant charge transfer with the first layer will facilitate the formation of polymorphs that have a large LUMO-LUMO overlap. Because the LUMO-LUMO coupling is a relevant ingredient for the electron mobilities of the compound,[8,69] it stands to reason that these polymorphs generally display superior properties. In our case, we can estimate the rate of interlayer charge transfer for the two systems by calculating the electronic coupling between LUMO orbitals with the Projection-Operator Diabatization method.[70,71] The details of the Electronic Coupling calculation can be found in the Supporting Information. The results are shown in Table 1.

One can see that MoM exhibits superior electronic coupling over MoG for all the systems we consider. For the single molecular dimer from Figure 4 the difference is very large, and although a part of this difference is due to the nature of the dimer model, as discussed in the Supporting Information, the trend is persistent for more complex systems, up to and including the full bilayer geometries found by our structure search.

*Table 1. Interlayer electronic couplings for MoM and MoG structures*

| **Electronic Coupling (eV/molecule)** | Molecule-on-Molecule | Molecule-on-Gap |
|---|---|---|
| Molecular Dimer | 0.320 | 0.106 |
| Bilayer on Graphene | 0.269 | 0.209 |
| Bilayer on Ag(111) | 0.372 | 0.277 |

This shows that the influence of the choice of the substrate is crucial for the performance of any device, and exemplifies that, even when we can examine the fortuitous case in which two different substrates would seem to induce the same geometry to the first layer, the influence beyond the first layer can be enough to drastically alter the geometry and, thus, the properties of the system.

In conclusion, we have studied the structure of the first two layers of benzoquinone on two different substrates. Employing first-principles calculations in combination with machine learning, we have found that for the first layer, similar structures are favorable for both substrates. For the second layer, two structures are very favorable for both systems, but their ranking is swapped for the two substrates. This difference in ranking is a consequence of the difference in LUMO-LUMO coupling for the two different structures in the second layer. Hereby, the MoM structure has a large coupling compared to the MoG structure. Without induced charge, the MoG is energetically more favorable compared to MoM. When charge is induced into the first molecular layer (as is the case for Ag) MoM becomes energetically stabilized due to the LUMO-LUMO coupling. This points to the fact that the two different structures induced by the two substrates would exhibit different vertical charge carrier mobilities. Our computational study therefore indicates that substrates which undergo notable charge transfer with the first layer are more likely to induce polymorphs with large(r) electronic coupling and, hence, charge-carrier mobilities.

**Computational Methods**

All calculations were performed using the FHI-aims package,[72] with the PBE[73] exchange-correlation functional and TS$^{surf}$ correction for long-range dispersion interactions.[74,75] All geometry optimization were conducted with the BFGS algorithm, converging the forces on each atom to a threshold of 0.01 eV/Å. For the Ag system, the first 6 layers of metal were kept fixed and the top 2 layers were allowed to relax. Graphene atoms were kept fixed.

For the Ag system, default tight basis sets were used for all chemical elements except Ag, for which a mixed-quality numerical basis set (see ref. [29] for details) was employed. The calculations were conducted with the repeated slab approach, using a dipole correction,[76] and setting a unit cell height of 80 Å. For finding all single-molecule local adsorption geometries on the surface, a three-step procedure was followed. First, a single molecule was relaxed at an arbitrary position on top of the substrate unit cell (consisting of a molecular monolayer on Ag(111), see Figure 1a). Secondly, we used a Gaussian Process Regression tool equivalent to the BOSS approach[46] to find all stationary points in the PES along three dimensions (translations along X and Y, rotation of the molecule around the axis perpendicular to the surface). Finally, all the geometries corresponding to these points were relaxed keeping all substrate atoms fixed. At this stage of the work, all calculations were executed on a 2x2 substrate cell, integrating in k-space on a grid of 3x3 points per primitive lattice direction and 1 k-point in the Z direction. Given the high computational cost of running geometry optimizations with these systems, the search for local adsorption geometries was conducted on a gas-phase monolayer substrate, in which Ag atoms were removed. The adsorption energy of the adsorption geometries was evaluated reintroducing the metal atoms for a single-point calculation. The SAMPLE approach uses these adsorption geometries as building blocks to assemble different configurations, placing them in all possible ways on a set of different unit cells (details in the Supporting Information). Among all configurations, a set of 250 was selected with experimental design employing the D-optimality criterion[77] on intermolecular interactions. Of these, 200 were used as training set, while the remaining 50 were used as test set. In addition, 961 gas-phase calculations, in which all substrate atoms were removed, were used to calculate priors for all intra-layer interaction energies. At this stage of the work, given the necessity to work with a wide variety of unit cells, the k-space integration was conducted on automatically generated generalized Monkhorst-Pack grids.[78] After training in the conditions described beforehand, SAMPLE predicts the adsorption energies of the test set with a root mean square error (RMSE) of 7 meV. Leave-one-out cross validation[79] (LOOCV) was also applied on the training set and gave a RMSE of 10 meV.

For the graphene system, the procedure is identical to the Ag case, unless specified otherwise. Default tight basis sets were used for all chemical elements. The lattice constant of graphene was converged to 2.46 Å, and a unit cell height of 85.2 Å was set. When searching for local adsorption geometries for the first benzoquinone layer, a 5 x 5 substrate cell was used, and the k-space integration was conducted on a grid of 6 x 6 points per primitive lattice direction and 1 k-point in the Z direction. For the SAMPLE prediction, 100 calculations were used, 60 as training set and 40 as test set, together with 1000 gasphase calculations, resulting in an RMSE of 5 meV on the test set and a LOOCV-RMSE of 7 meV.

When searching for local adsorption geometries for the second benzoquinone layer, the structure shown in Figure 1b was set as primitive substrate unit cell. A 2 x 2 substrate cell was used, and the k-space integration was conducted on a grid of 6 x 6 points. For the SAMPLE prediction, 250 calculations were used, 200 as training set and 50 as test set, together with 997 gasphase calculations, resulting in an RMSE of 14 meV on the test set and a LOOCV-RMSE of 23 meV. The calculation of electrostatic terms as shown in Figure 4b was conducted with a code designed for periodic systems (see ref [64] for details). To emulate a cluster system, the molecular dimer was placed in a 25x25x50 Å unit cell. Additional charge was added with a layer of point charges analogously to the CREST method.[80] The calculation of electronic coupling terms was performed with the Lowdin-orthogonalized[81] second version of the Projection-Operator Diabatization method POD2L,[70] which was recently demonstrated to yield very accurate results for organic molecules.[71]

For these calculations, FHI-Aims default light basis sets were used in place of the tight basis sets, as the former were found to be more numerically stable under the required block-diagonalization scheme.

**Supporting Information**

Details on the convergence of k-space integration grids; graphical representation of all local adsorption geometries; details about the SAMPLE prediction of the second layer on Ag(111) and of the first and second layer on graphene, inclusive of geometry optimizations; discussion of the best first layer polymorphs to be chosen as substrate for the second layer; analysis of adsorption-induced charge transfer and MODOS analysis for the two substrates; comparison of the dimer model to more complex models for the comparison of MoM and MoG, further discussion of the best configurations for the second layer of benzoquinone on graphene; details on the calculation of electronic coupling terms.

**Author Information**

The authors declare no competing financial interests.

**Acknowledgements**

We thank Christian Winkler for providing the code for the energy decomposition analysis and assisting in its usage. We acknowledge fruitful discussions with J. J. Cartus, A. Werkovits, B. Ramsauer and R. Steentjes. Funding through the START project of the Austrian Science Fund (FWF): Y1157-N36 is gratefully acknowledged. Computational results have been achieved using the Vienna Scientific Cluster (VSC).

# Supporting information to "Interfacial charge transfer influences thin-film polymorphism"

*Fabio Calcinelli[1], Andreas Jeindl[1], Lukas Hörmann[1], Simiam Ghan[2], Harald Oberhofer[2] and Oliver T. Hofmann[1]\**

[1]Institute of Solid State Physics, Graz University of Technology, 8010 Graz, Austria

[2]Chair for Theoretical Chemistry and Catalysis Research Center, Technical University Munich, 85748 Garching, Germany

*Corresponding author: **o.hofmann@tugraz.at**

# Contents

# 1. Convergence of k-grids

### BQ on Ag(111) 2nd Layer

For the calculation of local adsorption geometries, a grid of 6x6 k-points was used. Given that the calculations were run on a 2x2 cell, this corresponds to a 12x12 k-grid on the 1st layer unit cell. As the 1st layer unit cell has a surface corresponding to 12 Ag unit cells, arranged as a 3x4 grid, this value corresponds to a grid of 36x48 points, which is in excess compared to the value used in the reference. [1]

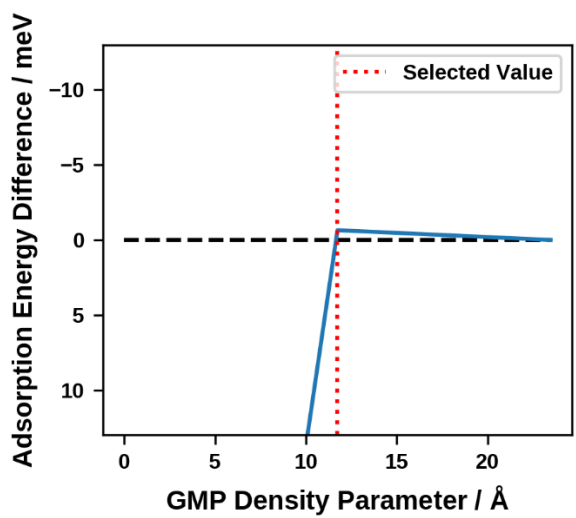

*Figure S 1. Convergence of adsorption energy with respect to GMP density parameter for the second layer of benzoquinone on Ag(111). The energy difference is computed with respect to the densest grid.*

For the calculations following the application of SAMPLE, automatically generated generalized Monkhorst-Pack (GMP) grids were used.[2] The density parameter (corresponding to the reciprocal of the maximum distance between k-points in reciprocal space) was converged to 11.74 Å, giving an error of less than 10 meV in adsorption energy for the most stable local adsorption geometry, as shown in figure S1.

## BQ on Graphene 1st Layer

For the calculation of local adsorption geometries, a grid of 30x30 k-points for a graphene primitive unit cell was selected (see figure S2) and adapted to the size of each supercell.

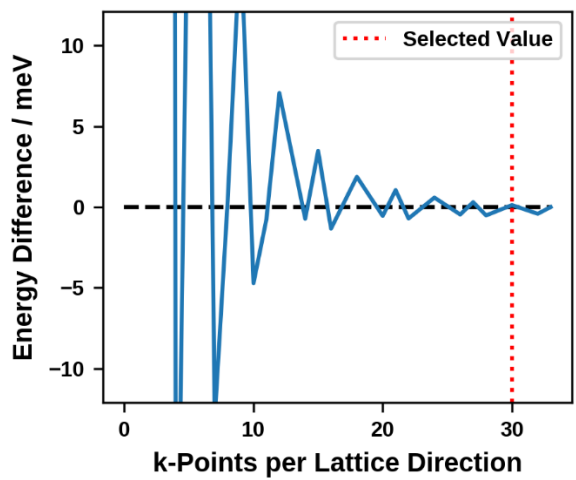

*Figure S2. Convergence of total energy with respect to number of k-Points for graphene. The energy difference is given with respect to the densest grid.*

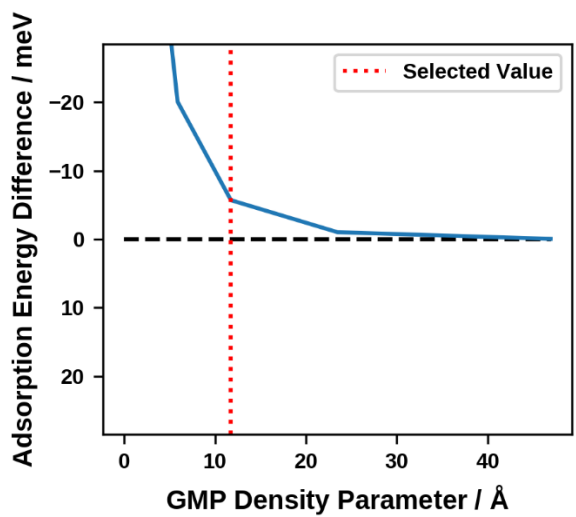

*Figure S 3. Convergence of adsorption energy with respect to GMP density parameter for benzoquinone on graphene. The energy difference is computed with respect to an overconverged system with 180x180 k-Points for primitive graphene unit cell.*

For the calculations following the application of SAMPLE, automatically generated generalized Monkhorst-Pack grids were used. The density parameter was converged to 11.74 Å, giving an error of less than 10 meV in adsorption energy for the most stable local adsorption geometry, as shown in figure S3.

## BQ on Graphene 2nd Layer

For the calculation of local adsorption geometries, a grid of 7x7 k-points was used. Given that the calculations were run on a 2x2 cell, this corresponds to a 14x14 k-grid on the substrate primitive unit cell. As the substrate primitive unit cell has a surface corresponding to 8 graphene primitive unit cell, this value is in excess compared to the value used for the 1st layer of BQ on graphene.

For the calculations following the application of SAMPLE, automatically generated generalized Monkhorst-Pack grids were used. The density parameter was converged as shown in Figure S4, and a value of 11.74 Å (the same used for the previous layer) was chosen for simplicity.

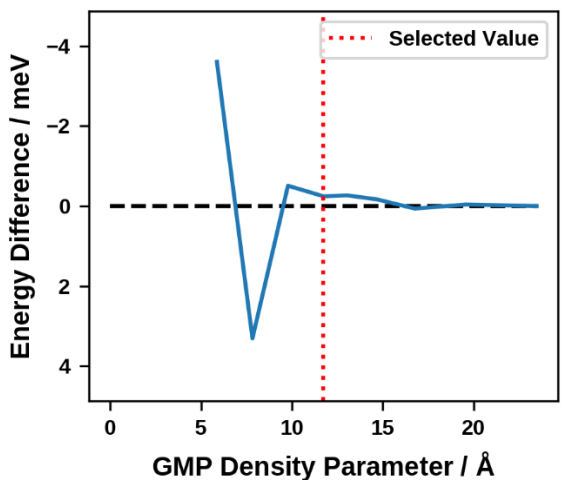

*Figure S 2. Convergence of adsorption energy with respect to GMP density parameter for the first layer of BQ on graphene. The energy difference is given with respect to the densest grid.*

## 2. Application of the SAMPLE approach for finding and ranking polymorphs

We use SAMPLE to find and rank all possible polymorphs for the systems we consider. In this section we include not only the application of the SAMPLE algorithm, but also the preliminary stage of finding local adsorption geometries (building blocks) and the geometry optimizations performed on the best structures.

### Finding Local Adsorption Geometries (building blocks)

As reported in the main text, the following procedure was applied to find local adsorption geometries (the building blocks for all configurations): first, a single molecule was relaxed at an arbitrary position on top of the substrate unit cell to find a suitable adsorption height; secondly, the BOSS approach[3] was employed to find all stationary points in the PES along three dimensions (translations along X and Y, rotation of the molecule around the axis perpendicular to the surface); finally, all the geometries corresponding to these points were relaxed while keeping all substrate atoms fixed. If two or more optimizations led to the same position, the redundant ones were eliminated. All local adsorption geometries for the 1st and 2nd layer of benzoquinone on graphene are reported in Figures S5 and S6, together with the relative adsorption energies.

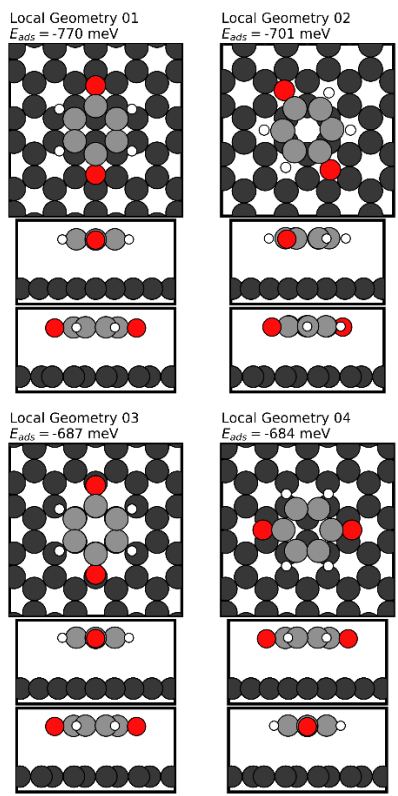

*Figure S5. Local adsorption geometries for the 1st layer of benzoquinone on graphene (top view and two orthogonal side views)*

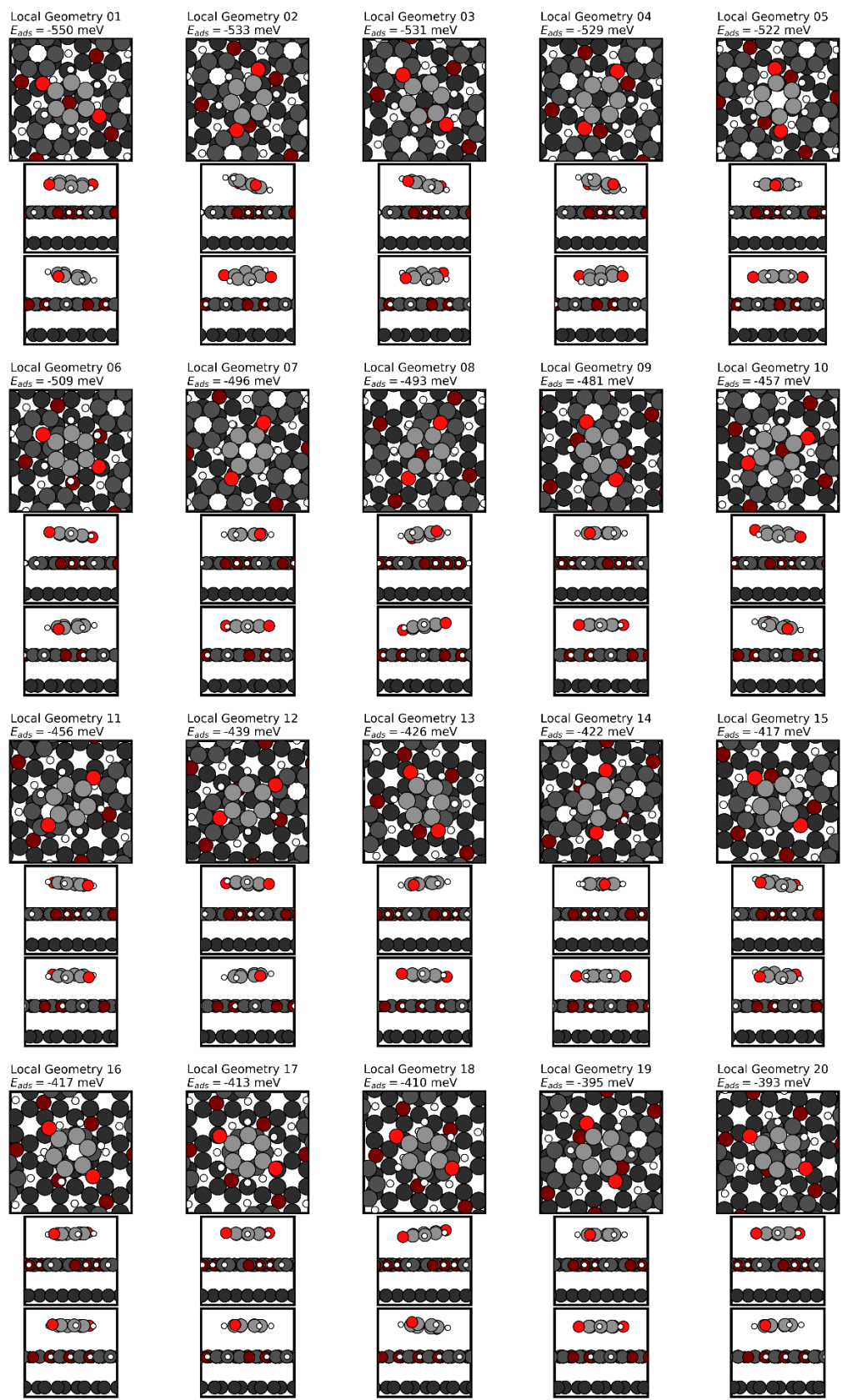

*Figure S6. Local adsorption geometries for the 2nd layer of benzoquinone on graphene (top view and two orthogonal side views)*

In the case of benzoquinone on Ag(111), local adsorption geometries for the second layer were found on a free-standing benzoquinone substrate, and the resulting geometries were then recalculated on the full metal substrate. All local adsorption geometries on the free-standing benzoquinone substrate are reported in Figure S7, while the variations in energy and ordering that result from the reintroduction of the Ag atoms are indicated in Table S1.

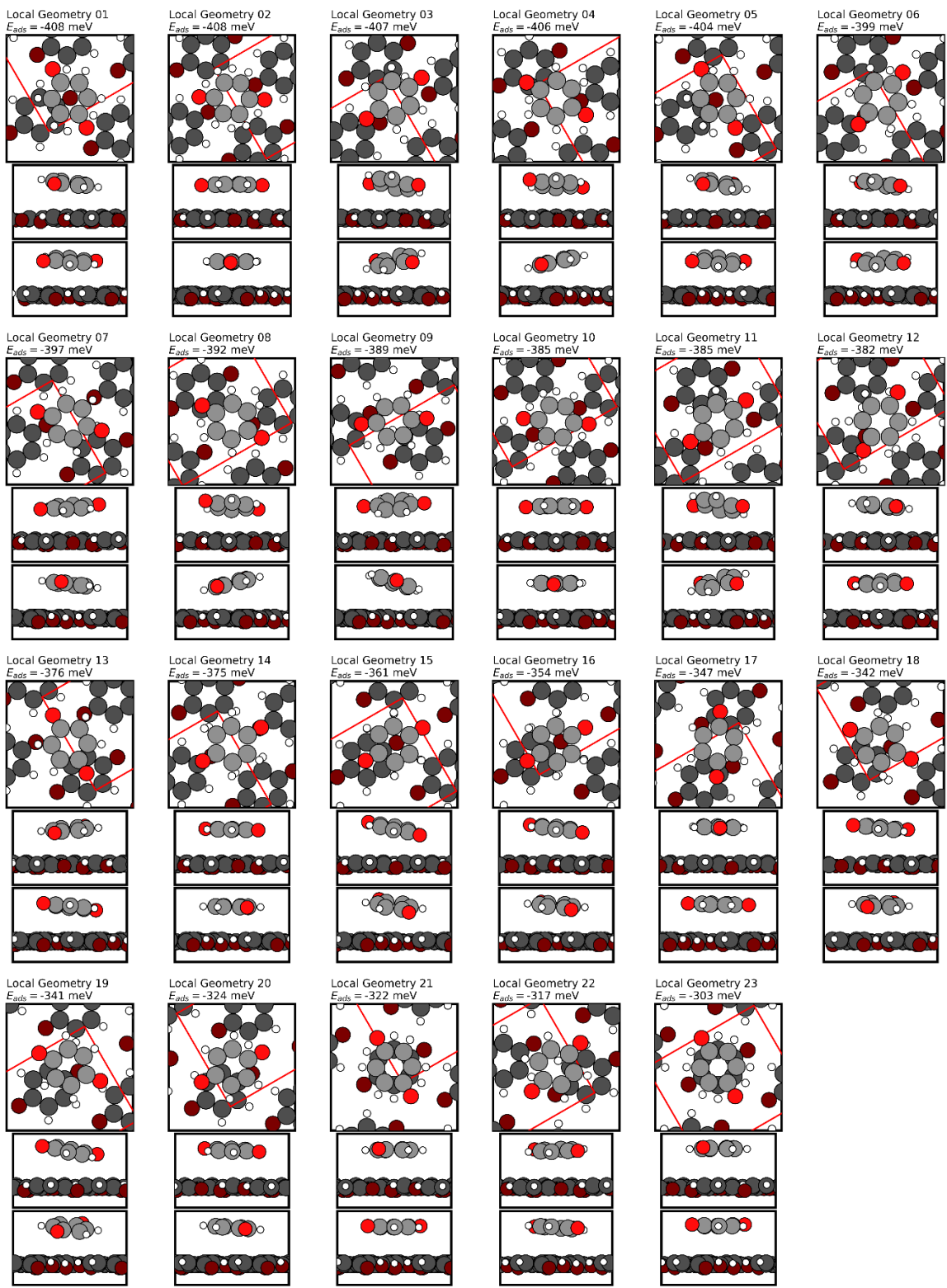

*Figure S7. Local adsorption geometries for the second layer of benzoquinone on Ag(111), as found on a free-standing monolayer of benzoquinone (top view and two orthogonal side views). The unit cell of the substrate is shown in red, allowing to distinguish the two non-equivalent molecules of the substrate.*

*Table S1. Adsorption energies of all local adsorption geometries of the 2nd layer of benzoquinone on graphene and original ranking and adsorption energy on the simplified free-standing-monolayer substrate*

| Local Adsorption Geometry | Adsorption Energy (meV) | Original Ranking on Free-Standing Monolayer Substrate | Original Adsorption Energy on Free-Standing Monolayer Substrate (meV) |
|---|---|---|---|
| 1 | -483 | 2 | -408 |
| 2 | -478 | 3 | -407 |
| 3 | -477 | 4 | -406 |
| 4 | -476 | 15 | -361 |
| 5 | -473 | 11 | -385 |
| 6 | -472 | 16 | -354 |
| 7 | -468 | 6 | -399 |
| 8 | -464 | 13 | -376 |
| 9 | -464 | 10 | -385 |
| 10 | -463 | 7 | -397 |
| 11 | -462 | 12 | -382 |
| 12 | -459 | 8 | -392 |
| 13 | -455 | 14 | -375 |
| 14 | -448 | 9 | -389 |
| 15 | -436 | 1 | -408 |
| 16 | -430 | 5 | -404 |
| 17 | -404 | 18 | -342 |
| 18 | -397 | 19 | -341 |
| 19 | -379 | 20 | -324 |
| 20 | -373 | 17 | -347 |
| 21 | -373 | 22 | -317 |
| 22 | -333 | 21 | -322 |
| 23 | -271 | 23 | -303 |

## Generating configurations with SAMPLE

SAMPLE produces a wide variety of configuration by producing a set of unit cells, and trying all combinations of local adsorption geometries and their symmetric equivalents that can be fit in each cell.[4] When executing this step of the SAMPLE approach, one must decide which cells to build, and how many molecules to try and fit in them. By using a big maximum cell size and a high maximum number of molecules, one would allow for the prediction of a larger number of configurations. This would of course allow for the possibility of finding new structures, but it must be noticed that the number of resulting configurations gets out of hand very rapidly, so one should always limit these parameters in order to obtain a reasonable number of structures. Moreover, one should modulate the maximum number of molecules to the size of the unit cell, to avoid producing a large number of mostly useless configurations with very low coverages. In addition to this, one must establish the distance threshold under which molecules are considered to be colliding, and the configuration containing them is discarded. This is defined for each possible combination of chemical elements, and must be chosen so that no configuration with strongly repulsive interactions is produced. The values of all these parameters, for the different systems on which SAMPLE was applied, are reported in table S2, together with the number of unit cells and configurations that were produced.

*Table S2. Parameters for the construction of configurations by SAMPLE and resulting numbers of cells and configurations.*

|  | 2nd Layer Benzoquinone on Ag(111) | 1st Layer Benzoquinone on Graphene | 2nd Layer Benzoquinone on Graphene |
|---|---|---|---|
| **Cell Areas (n. of primitive unit cells)** | 1 - 3 | 5 - 66 | 1 - 6 |
| **Number of molecules** | Area 1: 1 - 3<br>Area 2: 3 - 5<br>Area 3: 3 - 7 | Area 5 – 29: 1 – 3<br>Area 30 – 31: 3 – 4<br>Area 32 – 66: 3 | Area 1: 1 - 2<br>Area 2: 1 - 4<br>Area 3: 2 - 6<br>Area 4: 2 - 8<br>Area 5: 3 - 10<br>Area 6: 4 - 9 |
| **Distance Thresholds (Angstrom)** | HH: 1.500<br>OH: 1.500<br>OO: 2.200<br>CH: 2.000<br>CO: 2.400<br>CC: 3.000 | HH: 1.337<br>OH: 1.441<br>OO: 2.127 | HH: 1.500<br>OH: 1.500<br>OO: 2.200<br>CH: 2.000<br>CO: 2.400<br>CC: 3.000 |
| **Number of generated cells** | 7 | 1087 | 22 |
| **Number of generated configurations** | 83,044 | 26,518,330 | 349,483 |

## Rankings of all configurations

SAMPLE gives access to a prediction of the adsorption energies of all configurations. This allows to rank the configurations from the most stable to the least stable. In Figure S8 the rankings for all three SAMPLE runs are shown.

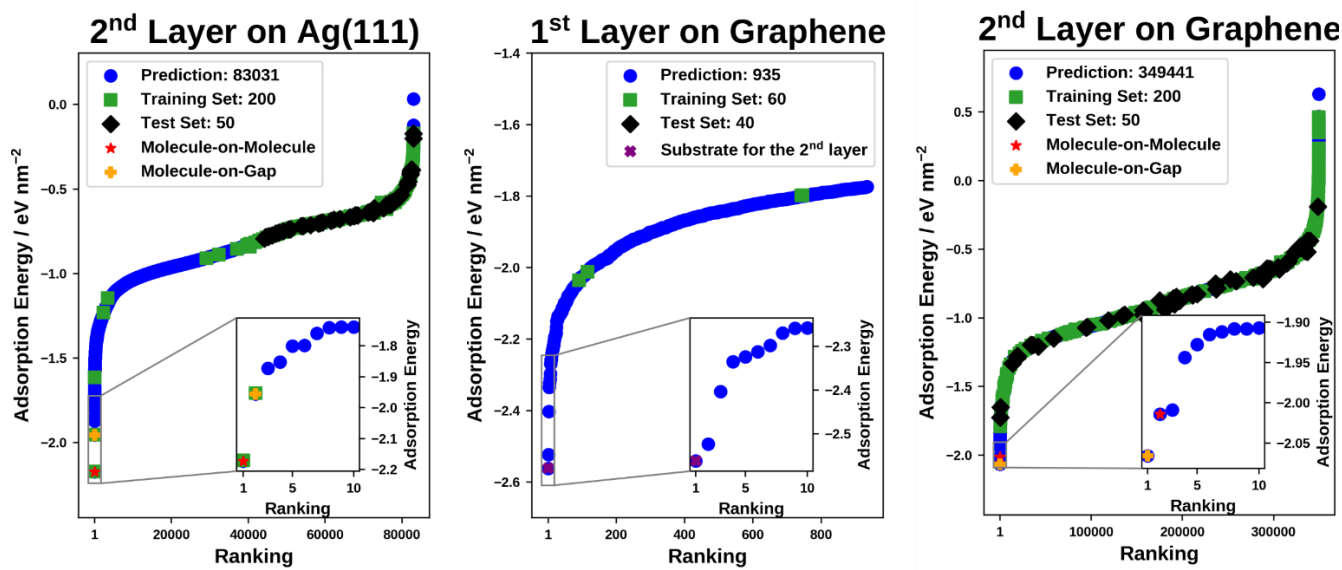

*Figure S8. Ranking of all configurations for all three systems. For the 1st layer of graphene, given the extremely large number of configurations, only a subset is shown. This is obtained by taking the best 1000 configurations and filtering out any configurations with reducible unit cells. The same duplicate removal procedure has also been applied to the top 1000 configurations of the other two systems, and this explains why the total number of plotted configurations is slightly smaller than the total number of constructed configurations reported in Table S2.*

### Geometry optimization of the best configurations

Once SAMPLE has allowed us to select the most stable configurations according to its energy model, the DFT energies for the 10 best configurations of each structures have been calculated. In addition, geometry optimizations are run allowing the newly formed layer of adsorbates to relax. Subsequently, also the first layer and – in the case of Ag(111) - the top layers of substrate are allowed to relax. This is necessary in the case of second-layer prediction, as the corrugated adsorption surface of the first layer combined with the generally weaker adsorption energies make it easier for the adsorbates to rearrange compared to their single-molecule adsorption geometry. A summary of these optimizations for the second layer of benzoquinone on Ag(111) is shown in Figure S9.

One can notice a very good agreement between the SAMPLE prediction and the single-point DFT calculations. On the other hand, the relaxation of the second layer of adsorbates produces big changes in energy, that fortunately do not change the ordering of the structures, except for structures 3 and 4. Structures 5-10 are very similar, and are fundamentally variations of structure 1 with some defects. On a positive note, one can notice that the re-relaxation of the 1st layer produces a small and uniform variation in energy, showing that the second layer of adsorbates does not influence the first layer strongly.

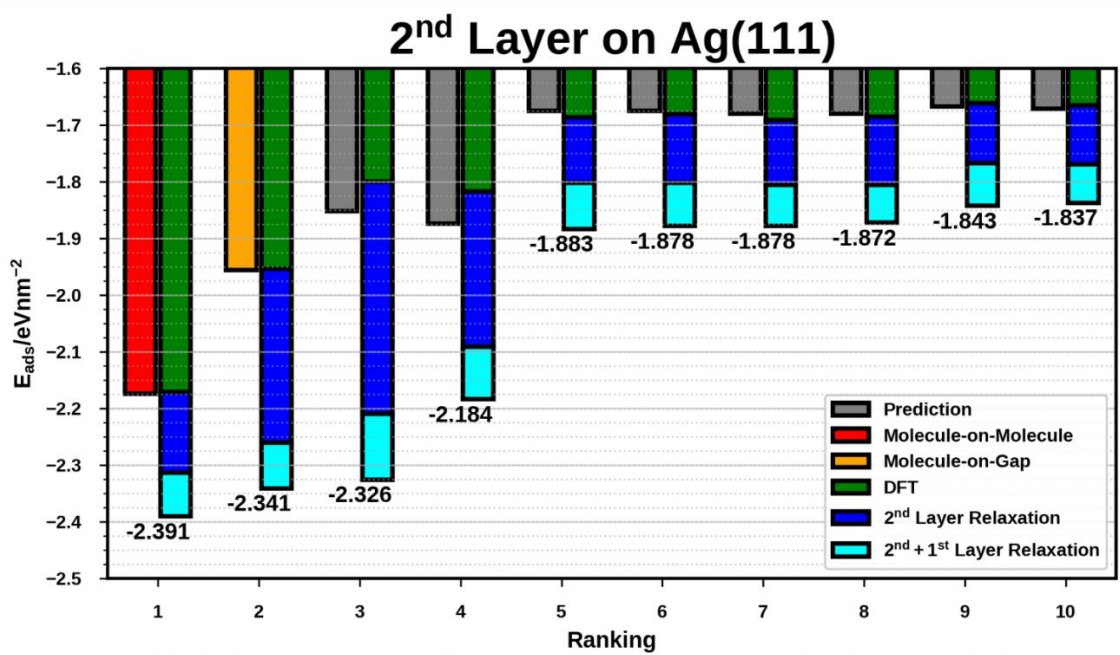

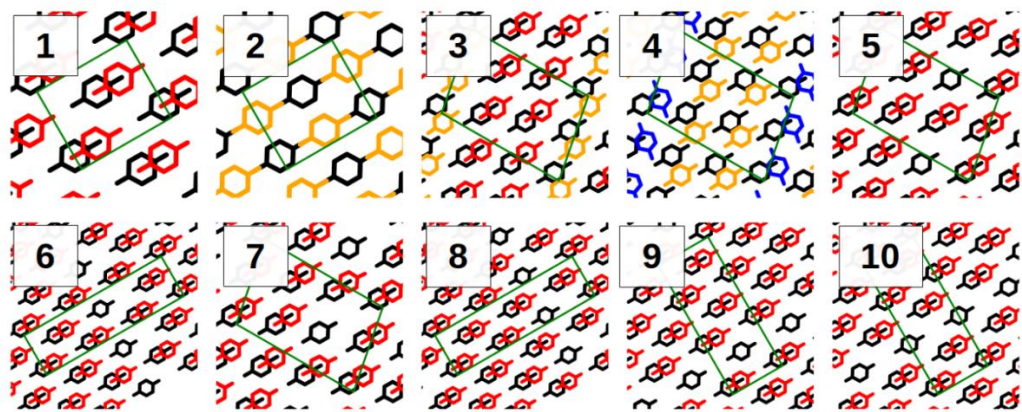

*Figure S9. Energies from predictions, single-point DFT calculations, and DFT geometry optimizations of the 10 best configurations of the 2nd layer of benzoquinone on Ag(111) and graphical representation of all 10 configurations*

A summary of optimization results for graphene is given in Figure S10. For this system, while the agreement between SAMPLE predictions and DFT data remains good, the geometry optimizations produce intense and irregular changes in energies, which result in several switches in ordering going from the prediction results to the post-optimization results. The relaxation of the first layer also produces stronger perturbations compared to Ag(111), as a consequence of the weaker interaction between substrate and 1$^{st}$ layer. In particular, we can see a set of structures with a very elongated unit cell arriving to the top 5 positions of the ranking. Of these, structures 1a to 1c are fundamentally equivalent, so 1b and 1c are discarded. Structures 3a, 3b and 3d are also basically identical to the single-cell configuration 3c and are therefore discarded. This gives us the top-5 ranking shown in Figure 2.

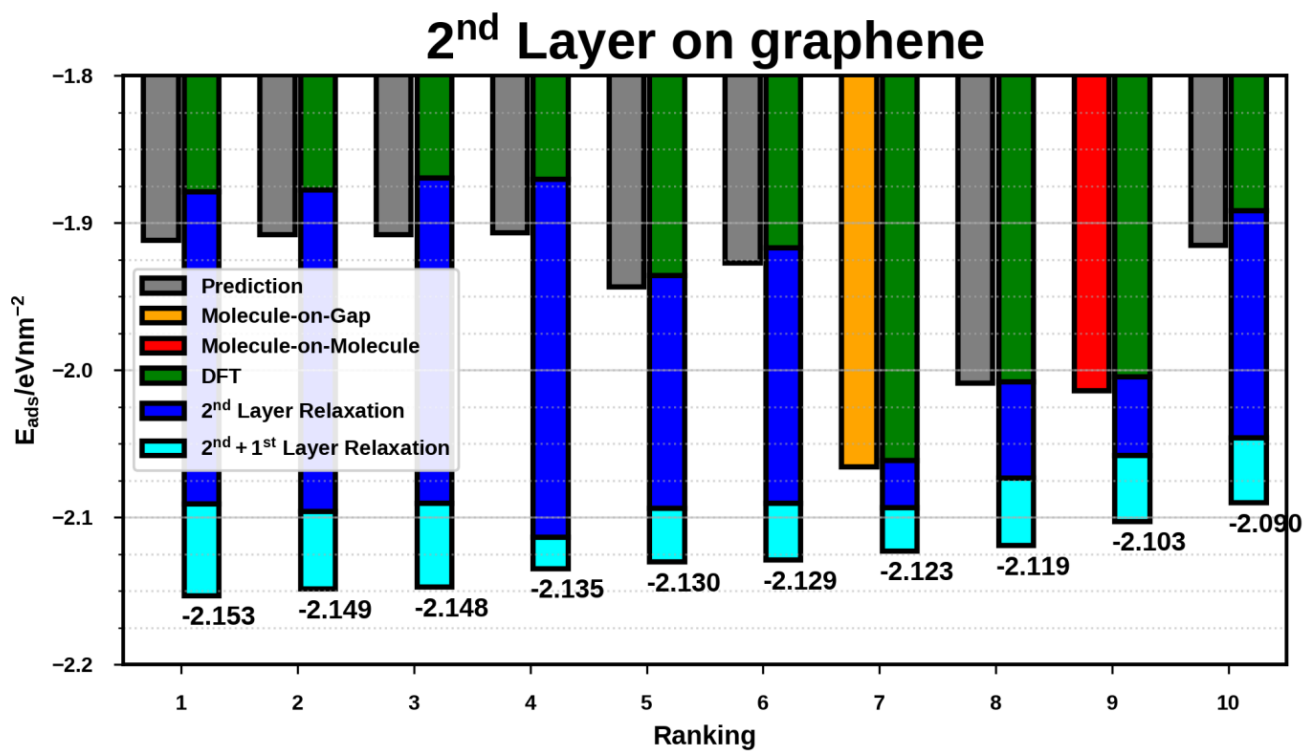

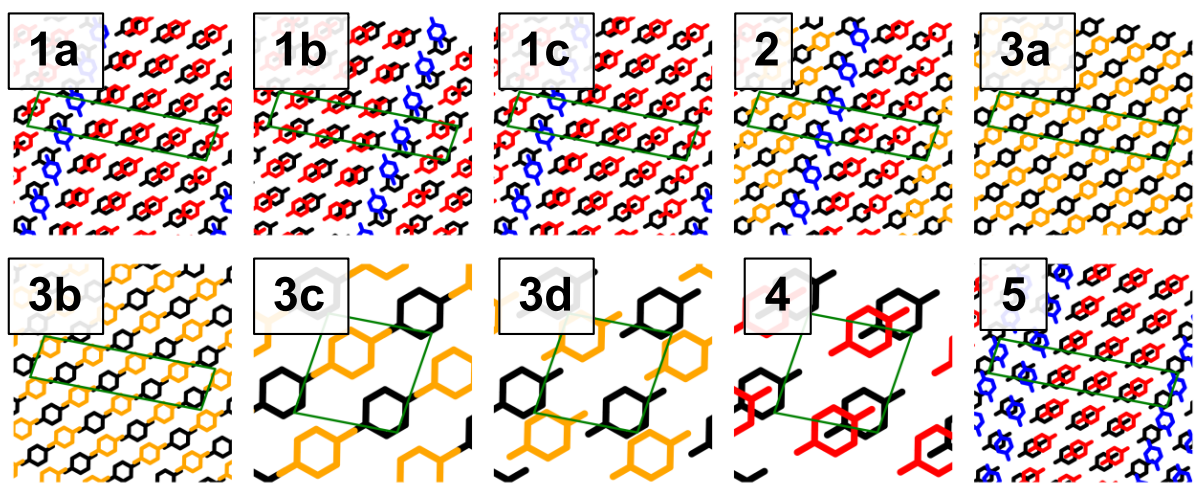

*Figure S10. Energies from predictions, single-point DFT calculations, and DFT geometry optimizations of the 10 best configurations of the 2nd layer of benzoquinone on graphene and graphical representation of all 10 configurations*

## 3. Selection of the first-layer polymorphs to be used as substrates for the growth of the second layer

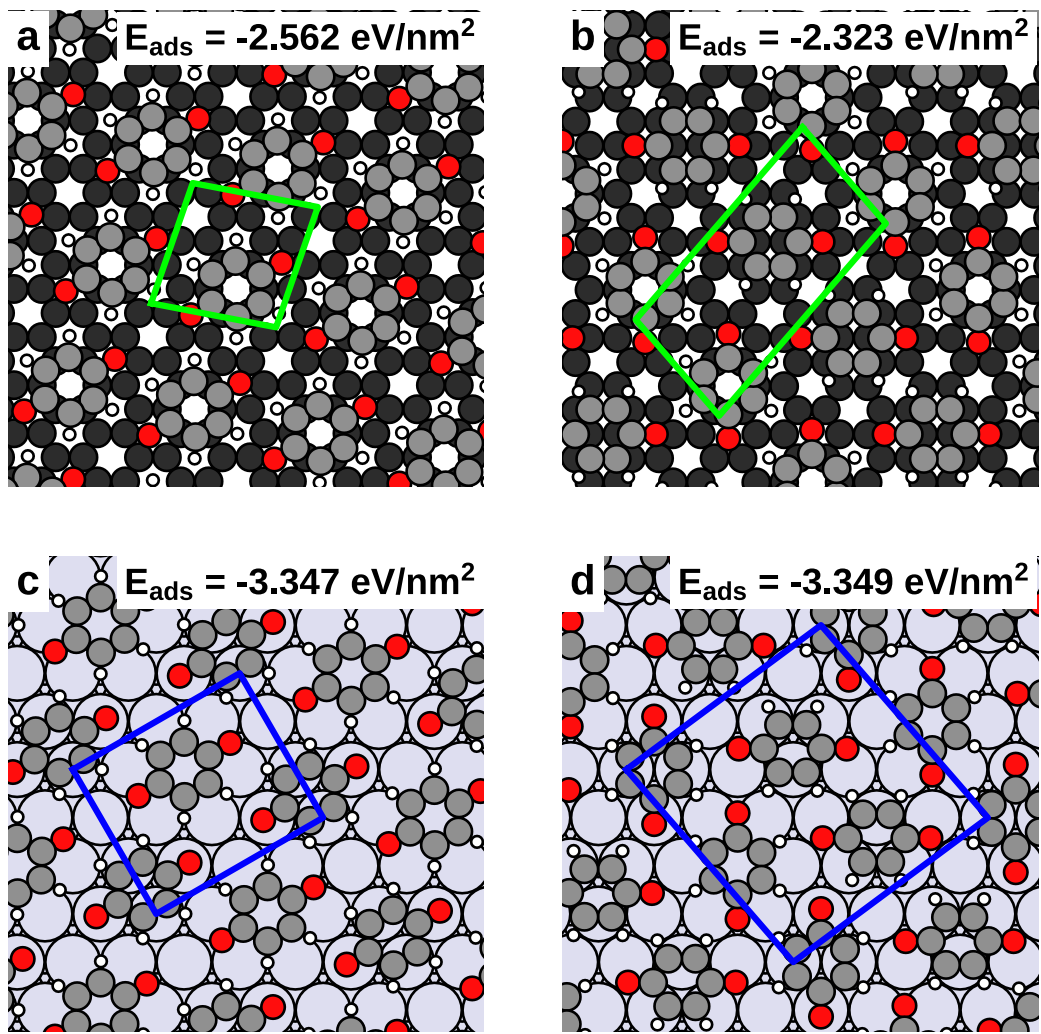

*Figure S11. 1st (a, selected as substrate for the second layer) and 5th (b) configuration of benzoquinone on graphene, according to adsorption energy per area. 2nd (c, selected as substrate for the second layer) and 1st (d) best configuration of benzoquinone on Ag(111).*

Applying the SAMPLE approach to the prediction of the second molecular layer requires choosing a first-layer polymorph as substrate. The most sensible criteria for the selection of the most stable closed-packed polymorph is the energy per area.[5] In the case of benzoquinone on graphene the most stable polymorph is the one we have selected as substrate for the second layer, with an adsorption energy per area of -2.562 eV/nm$^2$. It is shown in figure S11-a. This polymorph is also the best in terms of adsorption energy per molecule. The second best polymorph is an extremely similar structure, with the same molecular arrangement slightly shifted with respect to the substrate. After these two polymorphs, all other structures are less stable by more than 100 meV/nm$^2$. In the case of benzoquinone on Ag(111), a polymorph with an identical structure to the best structure on graphene is found, with an energy that is tied to the energetically best polymorphs within our prediction uncertainty (-3.347 eV/nm$^2$ $^{versus}$ -3.349 eV/nm$^2$). This structure, which we show in figure S11-c is the structure we have selected for the comparison of the two substrates.

## 4. Discussion of complex second-layer configurations of benzoquinone on graphene

As shown in Figure 2 and Figure S10, the 2$^{nd}$ layer of benzoquinone on graphene presented a few configurations more stable than Molecule on Molecule and Molecule on Gap, namely configurations 1 and 2 in Figure 2. While we have focused our discussion on simpler configurations, MoM and MoG, it is useful to discuss the properties of these more complex configurations. In particular, it should be understood what makes them so favorable, and why they appear for graphene and not for Ag(111).

First, it should be noticed that these structures share a fundamental feature: they are all constituted of a 5x1 unit cell, in which 4 molecules are aligned like MoM and MoG while the 5$^{th}$ molecule is rotated by 90°. The 4 aligned molecules are placed on the first molecular layer in positions similar to those of MoM and MoG, but each molecule is in a slightly shifted position, with respect to the first layer, compared to the previous one. The pattern is identical for configurations 1 and 2, the only difference being a small difference in alignment with respect to the first molecular layer. Consequently, we will now focus on configuration 1 to explain the cause of its stability, and the results will also apply to configuration 2.

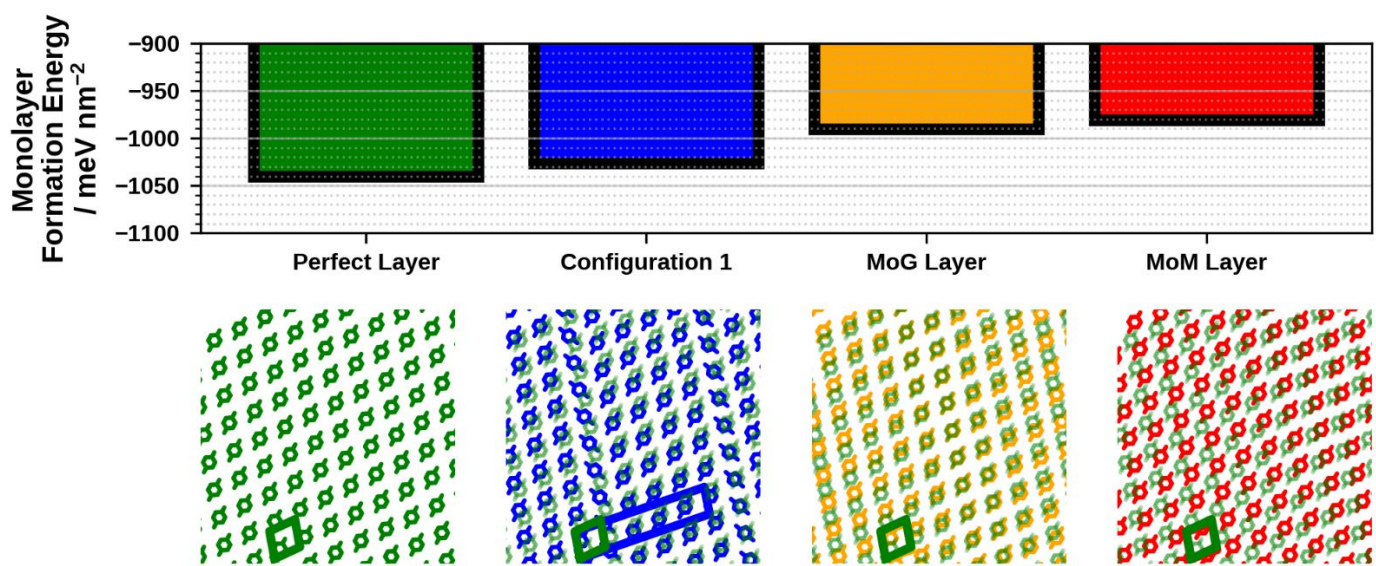

Figure S12. Comparison of monolayer formation energies for a perfect layer of benzoquinone (see main text), configuration 1 from figure S10, and the MoG and MoM configurations

To gain insight about the stability of configuration 1, we focus on intralayer interactions. To do this, we consider the energy of the configuration 1 layer as a free-standing layer in vacuum. The monolayer formation energy of this configuration is plotted in figure S12. For comparison, we have the monolayer formation energies of the MoM and MoG configurations, as well as the monolayer formation energy of a benzoquinone perfect layer. With perfect layer we indicate the geometry assumed by a benzoquinone sheet with the same geometry as MoG after a geometry optimization in which the unit cell vectors are allowed to relax, forming the geometry with the best intralayer interactions. We can see in figure S12 that the intralayer interaction energy of MoG and MoM are extremely similar, with MoG being around 10 meV more stable. This is well predictable, given that the two configurations are almost identical, except for their different alignment on the first layer, and for a slight tilting that MoM molecules adopt to fit on top of the first layer molecules. Configuration 1, on the other hand, is 40 meV more stable than MoG, and represents a middle point between MoM-MoG and the perfect layer. We can also see that, while the periodicity of MoG and MoG (which is the same as the periodicity of the first molecular layer) is incommensurate to the periodicity of the perfect layer, in the case of configuration 1 the 4 aligned molecules are almost perfectly congruent to the molecules of the perfect layer. In conclusion, we see that the arrangement of the

5 molecules of configuration 1 allows very favorable interlayer interactions, using the elongated 5x1 cell to create stripes of 4 molecules aligned at a very favorable angle, which is different from that of the first molecular layer and is very similar to that of a perfect benzoquinone layer. Therefore, we can conclude that it is a configuration that performs better than MoG and MoM because of its intralayer interactions and not because of a different mechanism of interactions with the previous layer. To understand why this configuration emerges in the case of graphene and not in the case of Ag(111), we must again notice how its cell is an elongated 5x1 cell, in which 5 molecules can be arranged in a row, so that 4 of them align at a very favorable angle, while the 5$^{th}$ molecule closes the gap and allows the structure to fit on the lattice of the first layer. If we now look at the unit cell of the Ag(111) first layer in Figure 1, we can see that it includes two molecules. Therefore, in order to replicate the 5-molecules patter of configuration 1, SAMPLE would need to assemble cells made of 5 primitive unit cells (10 molecules). As indicated in Table S2, this was not the case in our study, because extending the SAMPLE approach to cells of such dimensions on Ag(111) would pose an extremely high computational cost. Therefore, such structures were not generated by SAMPLE in the case of Ag(111).

## 5. Adsorption-induced charge-transfer and MODOS analysis of benzoquinone on Ag(111) and graphene

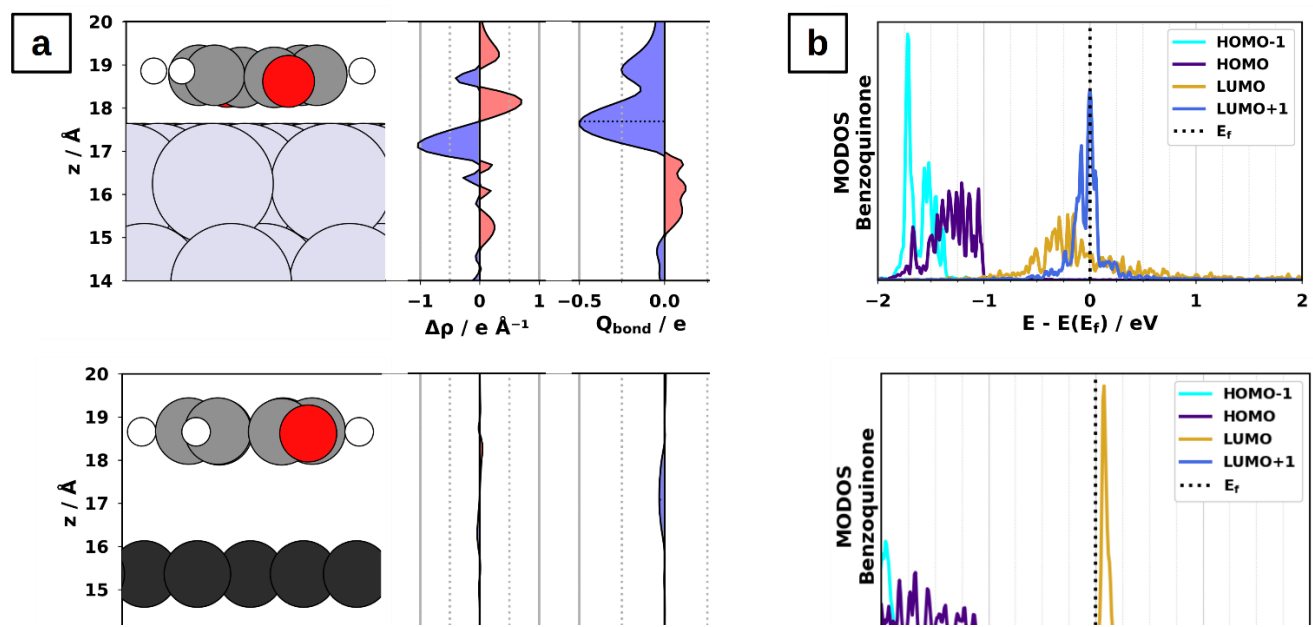

*Figure S13. Adsorption-induced charge transfer (a) and Molecular Orbital-projected Density Of States analysis of the first layer of benzoquinone on the two substrates.*

We conducted an analysis of the charge transfer resulting from the adsorption of the first layer of benzoquinone on Ag(111) and graphene. This was achieved by calculating the electron density of the isolated substrate and the electron density of the free-standing benzoquinone monolayer and subtracting them from the electron density of the combined system. The result of this operation, summed over each XY plane, is shown in Figure S13-a, together with the cumulative sum of such quantity along the z axis. The maximums of these sums are marked with a dotted line and correspond to the values reported in the text.

By conducing a Molecular Orbital-projected Density Of States analysis (MODOS) we derived the occupation numbers reported in Table 1. The results of this analysis are detailed in Figure S13-b, where one can see that the LUMO and LUMO+1 of the benzoquinone monolayer fall largely under the Fermi energy for Ag(111), but remain above it for graphene: this implies a large occupation number for Ag(111), and a very small occupation number for graphene.

## 6. Comparison of models of increasing complexity for reproducing the LUMO-LUMO overlap of different configurations

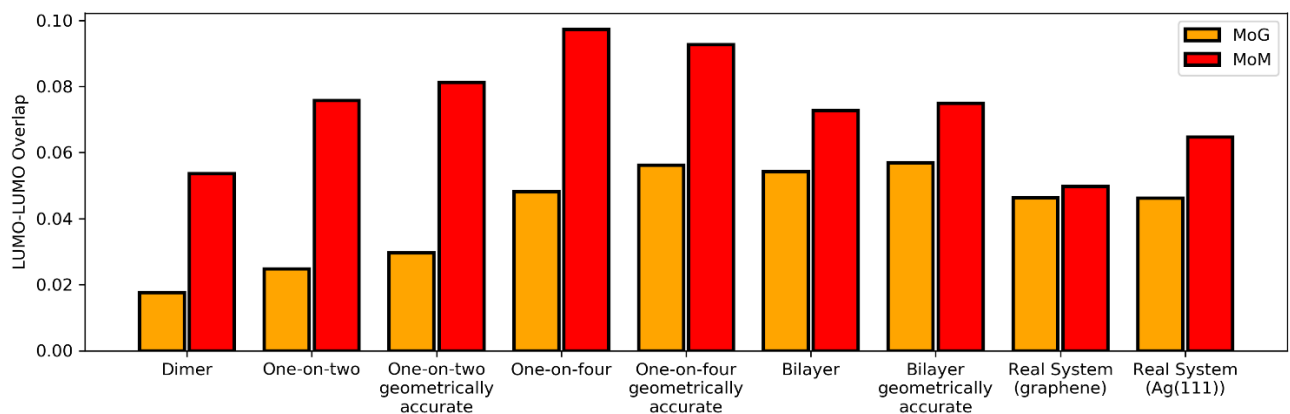

*Figure S14. LUMO-LUMO interlayer overlaps for systems of increasing complexity. Dimer: molecular dimer as presented in Figure 4; one-on-two: cluster with two neighboring molecules in the bottom layer (along the axis of displacement of the top molecule) and one molecule in the top layer; one-on-four: same as one-on-two, with two additional molecules in the bottom layer, at the sides of the first two; bilayer: same as dimer, but under periodic boundary conditions with the same unit cell as in the graphene bilayer; real systems: same structures as obtained with our structure search method, without substrates. "Geometrically accurate" indicates systems in which the intermolecular distance, the tilting of the top molecule and the bonding of the bottom molecule are adjusted to produce an exact replica of the molecules on Ag(111), specifically the strongly bent molecule and the molecule on top of it.*

In Figure 4 the values of LUMO-LUMO overlap for a benzoquinone dimer are shown and used to draw conclusions on the properties of the molecular bilayers found with our structure search method. It is easily noticed that a molecular dimer presents important differences from our real structures, and the suitability of such a dimer model for explaining the variation of overlap between our structures is not to be taken for granted. In particular, it can be noticed that while in Molecule-on-Molecule the top molecule is placed mainly on top of a single bottom-layer molecule, in Molecule-on-Gap the top molecule is placed between two adjacent bottom-layer molecules. As a consequence, a dimer model in which the top molecule can, independently of its position, only interact with one single bottom-layer molecule could be expected to fail in accurately describing the difference between the two configurations. To verify whether this is actually the case, a comparison of LUMO-LUMO overlap has been conducted on a model system in which the top molecules interact with two bottom-layer molecules. In addition, an analogous comparison has been performed on a series of model systems of increasing complexity, including the actual structures of the organic bilayers on graphene and Ag(111) found with our structure search approach. The results of this set of comparisons are shown in Figure S14. It can be observed that the overlap difference detected for the simple dimer model is among the biggest, and passing through more complex systems produces wide variations in the measurements. Overall, the difference always favors MoM. In conclusion, the fact that the magnitude of the difference depends on a variety of factors, with no specific feature dramatically improving the results with respect to the simple dimer model, and that the general trend is preserved, justifies our usage of the dimer model.

## 7. Calculation of electronic coupling terms

The electronic coupling terms presented in Table 1 have been obtained with the methodology described in ref 6. [6] For the molecular dimer, we calculated the coupling between the LUMOs of the two molecules.

For the bilayer on graphene, we calculated the coupling between the LUMOs of the two isolated monolayers at the gamma point.

For the bilayer on Ag(111), a unit cell of each layer contains two molecules. As a consequence, for each layer the molecular LUMOs combine to form two orbitals, LUMO and LUMO+1 of the isolated monolayer, which are almost perfectly degenerate in energy. We calculated the interlayer couplings between all 4 possible combinations of orbitals (LUMO-LUMO, LUMO-LUMO+1, LUMO+1-LUMO, LUMO+1-LUMO+1), summed the 4 values and divided by 2 to avoid double counting.